\documentclass[epj]{svjour}
\usepackage{graphics}
\begin{document}

\newcommand{\be}{\begin{equation}}
\newcommand{\nn}{\nonumber}
\newcommand{\ee}{\end{equation}}
\newcommand{\bea}{\begin{eqnarray}}
\newcommand{\eea}{\end{eqnarray}}
\newcommand{\wee}[2]{\mbox{$\frac{#1}{#2}$}}   
\newcommand{\unit}[1]{\,\mbox{#1}}
\newcommand{\degree}{\mbox{$^{\circ}$}}
\newcommand{\ltish}{\raisebox{-0.4ex}{$\,\stackrel{<}{\scriptstyle\sim}$}}
\newcommand{\vs}{{\em vs\/}}
\newcommand{\bin}[2]{\left(\begin{array}{c} #1 \\ #2\end{array}\right)}
\newcommand{\pred}{^{\mbox{\small{pred}}}}
\newcommand{\retr}{^{\mbox{\small{retr}}}}
\newcommand{\p}{_{\mbox{\small{p}}}}
\newcommand{\m}{_{\mbox{\small{m}}}}
\newcommand{\tr}{\mbox{Tr}}
\newcommand{\rs}[1]{_{\mbox{\tiny{#1}}}}	
\newcommand{\ru}[1]{^{\mbox{\small{#1}}}}

\title{Retrodictive states and two-photon quantum imaging}
\subtitle{}
\author{E.-K. Tan\inst{1}, John Jeffers\inst{1}, Stephen M. 
Barnett\inst{1} \and David T. Pegg\inst{2}
}                     
\institute{Department of Physics, University of Strathclyde, John 
Anderson Building, 107 Rottenrow, Glasgow G4 0NG, U.K.\and School of Science, 
Griffith University, Nathan, Brisbane, Queensland Q 111, Australia}
\date{Received: date / Revised version: date}
%
\abstract{We use retrodictive quantum theory to analyse two-photon quantum
imaging systems. The formalism is particularly suitable for calculating
conditional probability distributions.
} 
\titlerunning{Retrodictive two-photon quantum imaging}
\maketitle
\section{Introduction}
\label{intro}
Two-photon quantum imaging has been studied extensively for a number of
years, both experimentally and theoretically \cite{1}. The phenomenon
relies upon entanglement found in pairs of photons which are produced
by spontaneous parametric down-conversion in a $\chi^{(2)}$ crystal
\cite{mandel}. A
typical system is shown in figure 1.
\begin{figure}
\resizebox{0.75\columnwidth}{!}{%
  \includegraphics{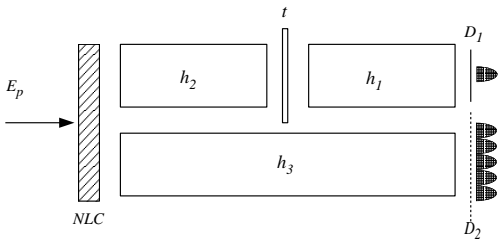}
}
\caption{A schematic 2-photon quantum imaging system. Arm 1 (upper)
contains a mask denoted by $t$ between the crystal and the detector. This
object can be imaged in arm 2.}
\end{figure}
A pump beam produces a pair of entangled photons in a type II
downconversion crystal. Due to the properties of both the pump field
$E_p$ and the nonlinear crystal (NLC) these photons are entangled in
energy and wavevector. The two photons within each photon pair are
emitted with polarisations orthogonal to one another, which enables
their separation by means of a polarising beam splitter. After
reflection or transmission at the beam splitter the photons travel on
their respective paths to be detected at spatially separated detection
systems. In arm 1 the photon usually propagates to a mask of some type
(with transmission function $t(x)$), whose image we wish to form, and
then after propagation it travels to the detector $D_1$, where it can
be recorded at a particular position in the transverse plane. In arm 2
there is not usually any mask, simply propagation to the detector
$D_2$. It is found that information about the object in arm 1 can be
found at the detector in arm 2, even though the two paths may be widely
separated, so that there is no chance that the photon in arm 2 could
have interacted with the object in arm 1. Of course this can only occur 
when the photon in arm 1 causes the detector $D_1$ to fire, so the 
information is conditional on the occurrence of this event.

A calculation of the spatially-dependent probability distribution for
joint photodetections at transverse position $x_1$ in arm 1 and position
$x_2$ in arm 2, $P(x_1,x_2)$, can give information about the object in
arm 1. The information is most directly obtained, however, from the conditional
probability distribution that there is a photodetection at $x_2$ given
that there is one at $x_1$, $P(x_2|x_1)$. This probability distribution 
is what is actually produced at $D_2$ in a multi-shot experiment, as the 
detections in arm 2 are only recorded if there is also one at detector $D_1$. 
It can be found from the
joint distribution using Bayes' theorem \cite{box},
\bea
\label{bayes}
P(x_2|x_1)=\frac{P(x_1,x_2)}{P(x_1)},
\eea
where of course, we assume that the arm 1 detection occurs within a
small neighbourhood of $x_1$ in the transverse plane, and take the
limit that the size of this neighbourhood tends to zero. There is
redundant information in the joint probability distribution
$P(x_1,x_2)$. It contains information, for example, about whether 
any photocounts are recorded by either detector due to the fact that the
nonlinear crystal normally does not produce any photon pairs within a 
detector integration time. The conditional probability disregards this 
extra information, as it only deals with cases where a photon is recorded 
at $x_1$. For this reason it would be better to calculate 
the conditional probability directly but, as we shall see later, there is
no way to do this in conventional predictive quantum mechanics. 

Klyshko \cite{klyshko} has suggested an advanced-wave interpretation
which has proved useful for the understanding of the results of such 
experiments. In essence the detected state in arm 1 is
thought of as evolving backwards through the system to the crystal,
where a conditioning of the 2-photon state takes place, forming a
1-photon wavefunction which evolves forward in arm 2 and is imaged at
the detector. The situation is similar to figure 2, which shows an
unfolded version of the system. The state evolves and the light is
thought of as propagating backwards through the system from the
detector to the crystal. Then the state evolves forward in time as the
light propagates from the crystal to the detection system in arm 2.
\begin{figure}
\resizebox{0.75\columnwidth}{!}{%
  \includegraphics{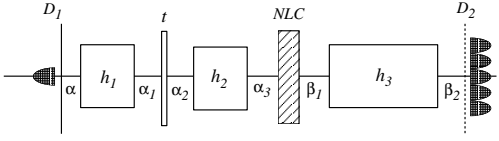}
}
\caption{Unfolded version of figure 1. The state evolves from the
detector on the left towards the detector on the right. The Greek letters
denote the spatial functions of the 1-photon states at each point in
the apparatus.}
\end{figure}

In this paper we utilise retrodictive quantum theory
\cite{bayesret,oldmaster,newmaster}, that is, quantum theory in which
the state of the system at any time between preparation and measurement
is assigned on the basis of a measurement performed on the final state
rather than the initially prepared state, to calculate directly
conditional probabilities, such as eq. (\ref{bayes}), for quantum
imaging systems. This approach has much to recommend it. Only the
required probability is calculated. The redundant information in the
joint distribution is not calculated because it is not useful. This is
the main advantage of  the retrodictive approach. Furthermore, it 
provides quantitative 
predictions based on a formal structure in which the reverse-time evolution 
of the measured state corresponds to Klyshko's advanced wave interpretation. 
This supports the Klyshko interpretation, provides a formal derivation of 
conditional probabilities, and thus makes the interpretation quantitative. 

The paper is organised as follows. In section \ref{retr} we describe
the basic features of retrodictive quantum theory. In the following
section we apply it to a general quantum imaging system such as in
figure 1. We then apply the theory to a specific example. Finally we
summarise our results and conclusions.

\section{Retrodictive quantum theory}
\label{retr}

Quantum theory is normally formulated in a predictive manner. It is
particularly useful if we wish to predict the outcomes of experiments
given particular initially prepared states. Thus it provides predictive
conditional probabilities. If we have a preparation device which prepares
states $\hat{\rho}_i$ with {\it a priori} probabilities $P(i)$, and we
measure these states with a device whose outputs $j$ are describable by a
probability operator measure (POM) with positive elements $\hat{\Pi}_j$
such that $\sum_j \hat{\Pi}_j = \hat{1}$ \cite{helstrom}, then the
predictive conditional probability that we obtain the result $j$ if the
prepared state was $\hat{\rho}_i$ is given by
\bea
\label{pcond}
P(j|i)=\tr \left( \hat{\rho}_i(t_m) \hat{\Pi}_j \right)
=\tr \left( \hat{U}(\tau) \hat{\rho}_i(t_p) \hat{U}^\dagger(\tau) 
\hat{\Pi}_j \right),
\eea
where $\hat{U}$ is the unitary evolution operator which evolves
the initially-prepared state from the preparation time $t_p$ to the
measurement time $t_m=t_p+\tau$.

If we do not know which state the preparation device prepared, but only
have access to the results of the measurement, then we require not the
predictive but the retrodictive conditional probability $P(i|j)$. This
is the probability that the state $\hat{\rho}_i$ was prepared given
that measurement result $j$ was recorded.  There are two ways in which
we can calculate this probability. Either we can calculate all possible
predictive conditional probabilities using predictive quantum
mechanics, and then use Bayes' theorem to find the retrodictive
probability, or we can use retrodictive quantum theory
\cite{bayesret,oldmaster,newmaster}. Retrodictive quantum theory is 
specifically designed to give the same results as predictive quantum 
theory combined with Bayes' theorem \cite{bayesret}. The Bayesian approach, 
however, is both more calculationally intensive and less elegant than 
using retrodictive quantum theory. 

In retrodictive quantum theory the state of a quantum system at any
time between preparation and measurement is the measured state evolved
backwards in time. At the preparation time the evolved measured state
collapses on to the preparation basis. It has been applied to both
closed systems, in which the time symmetry inherent in quantum theory
simplifies calculations greatly \cite{bayesret}, and to open systems,
where the retrodictive state evolves backwards in time according to a
retrodictive master equation analogous to the Lindblad master equation of
predictive quantum theory \cite{oldmaster,newmaster}. In closed systems
the retrodictive conditional probability that the prepared state was
$\hat{\rho}_i$ given that the later measurement result is $j$ is
\bea
\label{rcond}
\nn P(i|j)&=& \frac{\tr\left(P(i) \hat{\rho}_i \hat{\rho}_j(t_p)\right)}
{\sum_k \tr\left(P(k) \hat{\rho}_k \hat{\rho}_j(t_p)\right)}\\
&=& \frac{\tr\left(P(i) \hat{\rho}_i \hat{U}^\dagger(\tau)
\hat{\rho}_j(t_m)\hat{U}(\tau) \right)}
{\sum_k \tr\left(P(k) \hat{\rho}_k \hat{U}^\dagger(\tau)
\hat{\rho}_j(t_m)\hat{U}(\tau)\right)},
\eea 
where the retrodictive state $\hat{\rho}_j(t_m)= \hat{\Pi}_j/\tr
\hat{\Pi}_j$ is the normalised POM element corresponding to the
measurement result. This evolves backwards in time from the measurement
time to the preparation time, when it collapses on to one of the states 
which could have been prepared.

It is clear that there is an asymmetry in the forms of the predictive
and retrodictive conditional probabilities, equations (\ref{pcond}) and
(\ref{rcond}). This is not due to any inherent time-asymmetry in quantum
theory. Rather it is due to a choice in standard quantum theory to treat
the predictive conditional probability as fundamental, and normalise the
operators which describe prepared and measured states differently. Such a
choice is not necessary, and when preparation and measurement are treated
equally the predictive and retrodictive conditional probabilities take
on symmetric forms \cite{prepmeas,newmaster}.

\section{Retrodictive analysis}
\subsection{General theory}
\label{gentheory}

We now proceed to analyse the general system shown in figure 1. We wish
to calculate the conditional probability distribution of detection at a
general transverse position in arm 2 given a detection at a particular
transverse position in arm 1. We will formulate the theory in one
transverse dimension $x$. Extension to the whole transverse plane is
straightforward. 

In conventional quantum theory a fully predictive calculation
is performed based on the two-photon state produced by the crystal
evolved forward in time and space through both paths to form the joint
probability distribution of one detection in each path. We will calculate
only the conditional probability, performing a calculation which is part
retrodictive and part predictive in nature, in the spirit of the Klyshko
interpretation of such experiments. In order to simplify calculations
further we will dispense with the formal structure of density operators
and POM elements describing preparation and measurements, and simply
use prepared and detected states.

Suppose that a photon is registered by a detector centred at transverse 
position $x_1$ in arm 1. This is represented by the 1-photon state
\bea
\label{detstate}
|1_{x}\rangle_1 = 
\int dx \alpha(x) \hat{a}^\dagger(x)|0\rangle _1,
\eea
where $\alpha(x)$ is a normalised complex function of transverse
position centred on $x_1$, so that $\int dx |\alpha(x)|^2 = 1$. This
function gives the spatial profile of the detector. The continuous-mode
annihilation operator $\hat{a}(x)$ and the conjugate creation operator
obey the commutator \cite{contmode}
\bea
[\hat{a}(x), \hat{a}^\dagger(x^\prime)] = \delta(x-x^\prime).
\eea

The 1-photon retrodictive state can be evolved backwards in time from the
detection time. As it does so, we can follow the spatial profile back
through the apparatus to the point of preparation. This approach is typical of
Fourier optics \cite{fourier}. We denote the various
functions of $x$ at different points in the apparatus by $\alpha_1,
\alpha_2$ etc (fig. 2).  The first part of the propagation is the
propagation to the object. This is represented by convolution of the
spatial detector function $\alpha(x)$ with another function of $x$,
$h_1(x)$ to take account of the propagation. The state is still a 1-photon
state, but its spatial profile has become
\bea
\alpha_1(x) = \int dx^\prime \alpha(x^\prime) h_1(x-x^\prime).
\eea
The object which is to be imaged is accounted for by a simple transfer 
function $t(x)$, which is a spatially-varying complex function whose 
modulus is not greater than unity. Thus
\bea
\label{unnorm}
\alpha_2(x) = \alpha_1(x) t(x).
\eea
Note that a one-photon wavefunction with spatial function defined by
eq. (\ref{unnorm}) is not normalised. This is not a problem as we simply
normalise probabilities at the end of the calculation.  The next part
of the propagation is from the object to the crystal. Again this is
accounted for by convolution
\bea
\nn \alpha_3(x) &=& \int dx^\prime \alpha_2(x^\prime) h_2(x-x^\prime)\\
\nn &=& \hspace{-2mm}\int \hspace{-1mm}dx^\prime \alpha_1(x^\prime)
t(x^\prime) h_2(x-x^\prime)\\
 &=& \hspace{-2mm} \int \hspace{-1mm} dx^\prime \hspace{-2mm} \int
 \hspace{-1mm} dx^{\prime \prime}\alpha(x^{\prime \prime})
h_1(x^\prime-x^{\prime \prime}) t(x^\prime) h_2(x-x^\prime).
\eea
Thus the retrodictive state at the crystal is the 1-photon state of the
form defined by eq. (\ref{detstate}), but with spatial profile $\alpha$
replaced by the convolution $\alpha_3$.

We now condition the predictive state of the crystal using the 
retrodictive state from arm 1. The output of the crystal is assumed to 
be a 2-photon state of the form
\bea
|2_{x,x^\prime}\rangle_{1,2} = \int dx \int dx^\prime \beta(x,x^\prime) 
\hat{a}^\dagger(x) 
\hat{b}^\dagger(x^\prime)|0\rangle_{1,2},
\eea
where $\hat{b}^\dagger$ is the creation operator for arm 2. 
On conditioning this forms the one photon state in arm 2
\bea
\label{arm2state}
|1_x\rangle_2 = _1\hspace{-1mm}\langle
1_x|2_{x^\prime,x^{\prime \prime}}\rangle_{1,2} = \int dx \beta_1(x)
\hat{b}^\dagger(x)|0\rangle_2,
\eea
where 
\bea
\label{beta1x}
\beta_1(x) =\int dx^\prime \alpha_3^*(x^\prime) \beta(x^\prime,x).
\eea
It is clear that by conditioning the 2-photon crystal state with the
retrodictive state from the detector in arm 1 we produce a 1-photon state
in arm 2. In fact the combination of the detector in arm 1 and the crystal 
formally
constitute a quantum state preparation device \cite{prepmeas,newmaster}.
The complex conjugate in this function reflects the fact that the state
evolution in arm 1 has been backwards in time.

The final part of the calculation consists of propagation to the
detector in arm 2. This is again taken account of by convolution with
the propagation function $h_3(x)$. Thus
\bea
\beta_2(x) = \int dx^\prime \beta_1(x^\prime) h_3(x-x^\prime),
\eea
with the state given by eq. (\ref{arm2state}) with $\beta_2$ as the 
spatial profile.

The conditional detection probability distribution for obtaining a
detected photon at position $x_2$ in arm 2 given a detection at $x_1$
in arm 1, $P(x_2|x_1)$ is simply given by the squared modulus of the
final spatial profile, effectively a multiple spatial convolution of
all of the spatial functions
\bea
\label{genans}
P(x_2|x_1) = \frac{\left| \beta_2(x_2) \right|^2}{\int dx_2 
\left| \beta_2(x_2) \right|^2}.
\eea
Note that we now must divide by the integral of  the function in order 
to normalise the probability distribution. In principle we could have 
renormalised the 1- and 2-photon wavefunctions and this would have had 
the same effect.
\subsection{Example: Direct imaging of an object}

The utility of this general approach and of the formula derived in the
previous section can be illustrated by the following simple example
(see fig. 3). Suppose that there is a lens of focal length $f$ placed in
each arm of the system. The pair of photons produced by the crystal is
separated using a polarising beam splitter so that only one photon can
be counted at each detector. The detector in arm 1 is placed at the focal
length of the lens. The distance from the lens to the object to be imaged
is arbitrary. We assume, however, that the object, the beam splitter and
the crystal are all sufficiently close together that the small amount
of propagation here has no effect, so $h_2(x)=1$.  In arm 2 the crystal
is placed at the focal length of the lens, and a spatially-resolving
detector is placed after the lens. Any propagation after the lens then
has no effect on the results found at the detector in arm 2.
\begin{figure}
\resizebox{0.75\columnwidth}{!}{%
  \includegraphics{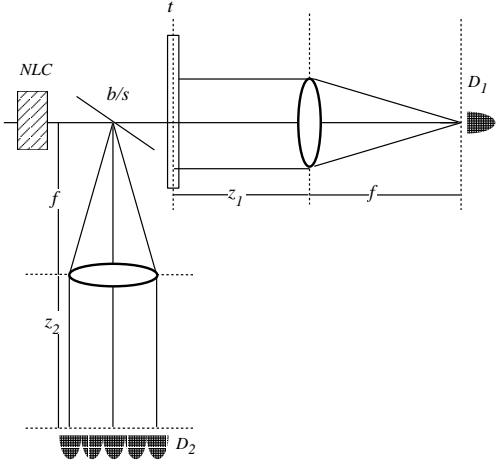}
}
\caption{The system considered in section 3.1. A pair of photons produced
by the crystal are separated at a polarising beam splitter (b/s) and detected
at separate detection systems.}
\end{figure}

The arm 1 detector resolution function will generally be of `top hat' 
form, but we will use a Gaussian 
\bea
\label{alphax}
\alpha(x)=\left(\frac{1}{\pi \sigma^2}\right)^{1/4}
e^{-(x-x_1)^2/(2\sigma^2)}
\eea
for ease of calculation. In any case when the spatial resolution of the
detector is either very good or very poor the exact form of the function
will not matter. It will be useful to write the state of the system in
Fourier space
\bea
|1_x\rangle =|1_{k_x}\rangle =\int dk_x \tilde{\alpha}(k_x) 
\hat{a}^\dagger(k_x)
|0\rangle,
\eea
where
\bea
\label{alphak}
\tilde{\alpha}(k_x) &=& \frac{1}{\sqrt{2\pi}}\int dx \alpha(x) 
e^{-i k_x x}\\
&=& \left(\frac{\sigma^2}{\pi}\right)^{1/4} e^{-k_x^2\sigma^2/2} 
e^{-i k_x x_1},\\
\hat{a}(k_x) &=& \frac{1}{\sqrt{2\pi}}\int dx \hat{a}(x) e^{-i k_x x}
\eea
are the transverse spatial Fourier transforms of the spatial function
and operator, and the vacuum state is now the state of no photons at any
transverse wavevector. For the Gaussian detector profile given above
the detected state has a transverse wavevector profile which is also
a Gaussian.

Propagation back to the lens corresponds to a modification of the 
transverse wavevector profile,
\bea
\tilde{\alpha}_1(k_x) = \left(\frac{\sigma^2}{\pi}\right)^{1/4} e^{ifk_z} 
e^{-k_x^2\sigma^2/2} e^{ik_x^2 f/k_z} e^{-i k_x x_1}.
\eea
The lens effectively takes the Fourier transform of this function, so
that components which propagate with different transverse wavevectors
between the detector and the lens all propagate with the same transverse
wavevector from the lens to the object, but with spatial profile given by
\bea
\label{alpha1x}
\nn \alpha_1(x) &=& \left(\frac{1}{\pi \sigma^2}\right)^{1/4}
\frac{e^{ifk_z}}{(1-2if/k_z\sigma^2)^{1/2}}\\
&\times&
\exp{[-(x-x_1)^2/(2\sigma^2(1-2if/k_z\sigma^2))]}.
\eea
As the spatial profile propagates unidirectionally, the distance
from the lens to the object is arbitrary, and we do not consider it.
After propagation back through the object the spatial profile becomes
$\alpha_2(x) = t(x) \alpha_1(x)$, with $\alpha_1(x)$ given by eq.
(\ref{alpha1x}). As was stated earlier, we assume that the crystal and
the polarising beam splitter are placed sufficiently close to the object
that the small amount of propagation involved makes no difference. Then
\bea
\label{alpha3x}
\alpha_3(x)=\alpha_2(x)= t(x) \alpha_1(x).
\eea 

The spatial profile of the 2-photon state is given by 
the functions
\bea
\label{betak}
\tilde{\beta}(k_x,k^\prime_x) &=& \sqrt{\frac{1}{2 \kappa^2}} 
\exp{[-(k_x+k^\prime_x)^2/2\kappa^2]},\\
\label{betax}
\beta(x,x^\prime) &=& \sqrt{\pi} \delta(x-x^\prime) 
e^{-x^{\prime 2}\kappa^2/2}.
\eea
The spread in transverse wavevector $\kappa$ of this function corresponds
to the spread in transverse wavevector of the Gaussian pump beam. Phase
matching then ensures that the photon pairs have wavevectors related
by eq. (\ref{betak}).  Projection of the back-propagated retrodictive
1-photon state onto this 2-photon state produced by the crystal produces a
1-photon state with profile given by eqs. (\ref{beta1x}), (\ref{alpha3x})
and (\ref{betax}),
\bea
\beta_1(x) = \sqrt{\pi} t^*(x) \alpha_1^*(x) e^{-x^2\kappa^2/2},
\eea
and $\tilde{\beta}_1(k_x)$ is found by Fourier transformation.
This state, prepared by conditioning a 2-photon predictive state with a
single photon retrodictive state, propagates forward in arm 2 from the
crystal to the lens placed at its focal length. This propagation again
corresponds to modification of $\tilde{\beta}_1(k_x)$ to form
\bea
\label{specans}
\tilde{\beta}_2(k_x) = \tilde{\beta}_1(k_x) e^{ifk_z} 
e^{ik_x^2 f/k_z}.
\eea 
The lens does the same as in arm 1, and effectively takes the Fourier
transform of the function, giving a probability distribution which
depends upon the transverse coordinate as in eq. (\ref{genans}). Again,
any further propagation from the lens to the detector causes no change
in the profile.

The result given in eq. (\ref{specans}) can be specialised for particular
arm 1 detector profiles. In particular, for a narrow profile, given by
the limit where $\sigma \rightarrow 0$. For a crystal which produces
a sufficiently broad spread of wavevectors, the transverse probability
distribution for detection in arm 2 takes on the form of $|t(x)|^2$. Thus
the image of an object in arm 1 is formed at the detection system in
arm 2, even though the photon in arm 2 never interacted with arm 1 at all.

The other extreme is given by a broad detector in arm 1. This gives the
marginal distribution $P(x)$ which will contain no spatial information
about the object in arm 1. The image is completely washed out by the
broad detector.

Other propagation/detection systems in arm 2 will give a different
profile. For example if the lens is placed a distance $2f$ from the
crystal, and the detectors are also a distance $2f$ from the lens then for
a `point' detector in arm 1 the probability distribution in arm 2 is the
squared modulus of the spatial fourier transform of the function $t(x)$.

\section{Conclusion}

In this paper we have used retrodictive quantum theory in a 2-photon
quantum imaging system to calculate the conditional probability
distribution for detection of a photon at a particular transverse position
in one arm, given a detection at another particular position in the other
arm.  The retrodictive state evolves backwards in time from the detection
in one arm to the nonlinear crystal, where it conditions the state of the
second photon.  This conditioned state then evolves forward in time in the
other arm and forms the probability distribution. The approach formalises
the interpretation of Klysko. We have calculated the general probability
distribution as a convolution of all of the transverse spatial effects
in both arm 1 and arm 2, and illustrated this with a specific example.

The advantage of the retrodictive approach over conventional predictive
quantum mechanics is that only the required probability distribution is
calculated. Much of the information in the full predictive probability
distribution for obtaining two detections at two distinct points in the
transverse plane, one in each arm is unnecessary.

In conventional quantum theory the 2-photon state is prepared by the
crystal, which forms a state preparation device. The two detectors,
one in each arm are measurement devices. The part-retrodictive,
part-predictive approach that we have described here represents the
system differently. The crystal together with the detector in arm 1
formally constitute a composite 1-photon state preparation device which
prepares a photon in arm 2 whose properties are determined nonlocally
in time both by physical processes in the crystal, and the later details
of the propagation to the detector in arm 1.

The authors would like to thank the following bodies for financial
assistance: the Overseas Research Students Awards Scheme, the University
of Strathclyde, the UK Engineering and Physical Sciences Research
Council, the European Commission (project QUANTIM (IST-2000-26019)),
and the Australian Research Council.

\end{document}